# Performance Analysis of Lifetime Maximizing Trees for Data Aggregation in Wireless Sensor Networks

Deepali Virmani , Satbir Jain

**Abstract- In this paper the performance of the proposed DLMT and CLMT algorithms are compared. These proposed algorithms tend to extend the node lifetime. Decentralized lifetime maximizing tree(DLMT) features in nodes with higher energy to be chosen as data aggregating parents while Centralized Lifetime Maximizing Tree(CLMT) features with the identification of the bottleneck node to collect data in a central manner among given set of nodes. Simulation results show that the functional lifetime is enhanced by 147% when data is aggregated via DLMT and by 139% when data is aggregated via CLMT. Proposed DLMT algorithm has shown 13% additional lifetime saving without increasing the delay. Packet delivery ratio show remarkable increase when the tree depth is considered in these tree structures.**

**Keywords- Wireless Sensor Networks, Data aggregation, E-Span, Network lifetime, Branch energy**

## I. INTRODUCTION

Sensor networks deploy heterogeneous collections of sensors capable of observing and reporting on various dynamic properties of their surroundings in a time sensitive manner.

Such systems suffer bandwidth, energy, and throughput constraints that limit the quantity of information transferred from end-to-end. These factors coupled with unpredictable traffic patterns and dynamic network topologies make the task of designing optimal protocols for such networks difficult. Mechanisms to perform data-centric aggregation utilizing application-specific knowledge provide a means to augmenting throughput, bandwidth and energy utilization. Data in such networks are not directly transmitted to interested users upon event detection. Instead, they are aggregated with neighboring sources locally to remove any redundancy and produce a more concrete reading. In this paper, comparison of construction of a data aggregation trees among any given set of source nodes is focused. The trees have a dedicated root for which the data from various sources are gathered and they are structured in such a way as to preserve the functional node lifetime of the event sources subject to the condition that they are constantly transmitting. The functional node lifetime is defined as the time till a node runs out of its energy. References [1-4] suggest that extending the node lifetime is equivalent to increasing the amount of information gathered by the tree root when the data rate is not time-varying.

Let

$BE_{a,b,N}$ : Energy of branch N leafed at node a and rooted at node b, N ε Pa,b .

$Et_x$ : Energy of tree rooted at node x.

Branch energy and Tree energy is calculated as:

$$BE_{a,b,N} \quad \min_{x\ N,\ x\ N} \{ e_x \} \qquad (1)$$

$$Et_x \quad \min_{y\ x,\ y\ a} \{ e_y \} \qquad (2)$$

Data reports from these sensors are clock- driven upon event detection. Furthermore, they are aggregated along their ways to be collected at the tree root and periodically sent to the sinks. To prevent data loss, the tree is periodically scanned and any broken link should be repaired whenever necessary. Such trees allow all raw data reports to be aggregated along the way to a single processing point. Only relevant information is extracted before transmitting it to any distant sink. Therefore, the converged tree construction becomes one of the fundamental issues for aggregation in WSNs. In fact, not all the trees are ideal for aggregation inside the event region. Since energy is usually scarce in WSNs, it would be most power-efficient if these sources can provide data to the sinks for the longest possible time. A tree that can survive for longer duration thus naturally becomes the best choice.

E-Span[5] improves the design of tree construction by assigning root to be the highest energy node. Such arrangement provides root with the maximum amount of energy resources for its additional duty in coordinating then route to distant sinks. However, there is still a high chance of assigning low-energy nodes to be the data aggregating agents for the other sources. To shorten the time and minimize the energy cost to tree reconstructions, and hence preserve the functional lifetime of all sources, a Decentralized lifetime Maximizing Tree construction algorithm (DLMT) and a Centralized Lifetime Maximizing Tree construction algorithm(CLMT) is proposed.

## II. Decentralized Lifetime-Maximizing Tree Construction Algorithm (DLMT)

DLMT[6] features in such a way that nodes with higher energy are preferably chosen as data aggregating parents whenever possible, so that the time to refresh this tree is extended and therefore less energy are involved in the tree maintenance. In addition, by constructing the tree in such a way, the protocol is able to lower the amount of data lost due to broken tree links before the tree reconstructions. Another attractive feature of the protocol is that the tree is most-likely to be centered in the middle of the event area, thereby reducing the delay during data collection. The goal is to construct a tree spanning all these sources and select an appropriate root for data collection, in a distributed way, such that the energy of the tree is maximized. The approach of exploring the highest -energy branch from each source to a root, by first assuming that every source node is a root, using a method similar to Reverse-Path Forwarding (RPF) [7] is used. This generates a total of N unique trees with each being rooted at a distinct source node. Comparison of the energy of these trees is done and the one with the highest tree energy for data collection /aggregation is employed.

## III. Centralized Lifetime Maximizing Tree Construction Algorithm (CLMT)

With CLMT[8] we assume that the complete knowledge of the event region including the connectivity and residual energy of all the source nodes is known prior to the start of this construction. The simple way to obtain CLMT is to directly run an extensive search at each node and then compare their tree energies. This method is very simple but has the scalability problem, when the network starts to grow or becomes dense the search becomes wider and takes a lot of time as well as the number of comparisons increase. So we tackle this issue with a complete different approach.

CLMT requires a root (initially unknown) to collect data from every other node via routes with the highest branch energy subject to condition that loop is not created. CLMT construction algorithm identifies the node that is causing a bottleneck to the set of connectivity provided by various event sources. Such arrangement extends the time to refresh the tree and lowers the amount of data loss due to broken tree link before the tree reconstructions. Thus tree also minimizes the delay and maximizes the lifetime of the source events by using minimum energy node as data collection node.

## IV. Simulation parameters

We implemented our tree construction modules on top of Forwarded Diffusion in the J-Sim network simulator (the J-Sim comes with diffusion support). In all of our experiments, a square sensor field with each side measuring X meters is being considered. A number of N identical nodes, ranging from 50 to 300 in the increment of 50, are randomly deployed in this sensor field such that the average node density is kept at $\lambda = 55/1652$ nodes per meter square, a parameter which we borrowed from Forwarded Diffusion [10, 11]. Furthermore, there are five sinks randomly deployed in the field and sources are randomly chosen among the nodes, subject to the conditions that SR=10% of N and the sources have to be interconnected to each other (to model a single stimulus). Each node is assumed to have a radio range of 45 meters. We considered an event-driven data sensor network throughout all our experiments. To model the periodic transmissions, each source generates random data reports of size fixed at 138 bytes in constant intervals of DR = 1 packet/second. To introduce some randomness, data start to be generated only after a time randomly chosen between t = 0 to 5 seconds. The data are collected at the root, if they exist, and are sent to the sinks. An application that computes the average of reports generated by various event sources is employed to model the aggregation behaviors. During data collection, sensors have the abilities to perform Fixed Application Independent Data Aggregation (FX-AIDA) of packets enrooted to the root. Specifically, we mean that each sensor can combine the reports received with itself into a single packet containing the average of all the gathered reports. We assign each source with an initial energy that is randomly chosen between 12 to 18 Joules in order to keep the total simulation time at a reasonable limit. In all of our experiments, all other nodes are given an initial energy that is greater than that of any event source such that their absence in the network, due to energy depletion, does not affect the functionalities of any participating sources during data collection. Lastly, the idle time power, receive time power and transmit power dissipation are set at 40, 400 and 680 mW respectively. We assume a negligible energy cost to process and aggregate incoming data reports. To trace the energy, an application that logs the residual energy of each node in constant intervals of 550 ms is employed. The J-Sim simulator implements a 1.6 Mbps 802.11 MAC layer. Since Forwarded Diffusion is chosen as our routing platform, we also adopt a range of Diffusion-related parameters listed in table 1 and simulation parameters listed in table 2 have been used in all of our experiments.

Table 1 : Diffusion-related parameters

| | |
|---|---|
| Packet Size = 86 bytes | Delay = 5 sec |
| Data packet = 138 bytes | Data Delay = 28 sec |

Table 2: Parameters used in simulation models.

| Parameters | Symbol | Values |
|---|---|---|
| Average Node Density | Λ | 55 / 1652 |
| Number of Nodes | N | 50,100,150,200 ,250,300 |
| Number of Sinks | S | 5 |
| Number of Sources | SR | 10 % of N |
| Network Width | X | ( N/ λ )0.5 |
| Node Energy | En | Variable |
| Data Rate | DR | 1 Packet/Sec |
| Protocol Timeframe | T | 28 sec |
| E-Span Control Size | SE-Span | 96 bytes |
| DLMT Control size | SDLMT | Variable |
| DLMT Hello Size | HDLMT | 63 bytes |
| Data Packet Size | – | 138 bytes |
| Radio Range | – | 45 m |
| Idle Time Power | – | 40 mW |
| Receive Time Power | – | 400 mW |
| Transmit Power | – | 680 mW |
| MAC bandwidth | – | 1.63 Mbps |
| Energy Log period | – | 550 ms |

*A. Performance Metrics for Simulations*

1) Average per Source Control ( ASC ) : ASC computes the amount of control cost in bytes for each source involved in construction and Maintenance of the data aggregation tree throughout a simulation run. It is calculated using equation (1).

$$ASC = \sum_{i=1}^{p} \frac{T_i}{p_u} l\, C \sum_{i=0}^{K-1} CP_t(i)\, CP_r(i+1) \quad (1)$$

δ → System wide discounting parameter $T_i$
→ Simulation Time
l → Total cost dependent on total messages exchanged in the interaction
C → cost of single message exchange
$CP_t(i)$ → Transmission cost of node i
$CP_r(i+1)$ → Receiving node of node i

$$CP_t(i) = \frac{\| d_i , d_{i+1} \|^n}{S_{i,i+1}} \quad (2)$$

$d_{i,i+1}$ → Cost due to distance between two nodes i and i+1

$S_{i,i+1}$ → Interference cost between two nodes i and i+1

2) Average Tree Depth ( ATD ) : ATD measures the average distance in number of hops between an event source and its tree root. It is calculated using equation (3).

$$ATD \qquad \frac{H^2 \quad h_1^2 \quad h_2^2}{2 \quad H \quad h_1} \tag{3}$$

H → Distance from the Sink/Root node to Source node

$h_1$ → hops of shortest path from leaf node to Sink/Root node

$h_2$ → hops of shortest path from leaf node to Source node

3) Average Dissipated Energy( ADE ) : ADE measures the average amount of energy consumed throughout the entire simulation. This metric computes the average work done in delivering periodic data to the sink/ Root over a simulation run. It is calculate using equation (4).

$$E_{cons} \qquad E_{cir} \quad KE_{amp} \quad d_{i,j} \quad 2 \quad K \tag{4}$$

$E_{cons}$ → Total Energy consumed

$E_{cir}$ → Energy consumed by the sensor node to run the circuit

$E_{amp}$ → Energy consumed by the amplifier

$d_{i,j}$ → Distance between node i and node j

K → Total number of data packets transmitted

4) Average Node Lifetime (ANLT ) : ANLT measures the time at which a source runs out of its available energy resource. This metric is required to determine how much additional time that each source can suffice by collecting data via the proposed tree structure. It is calculated using equation (5).

$$T^i_r \qquad \frac{E^l_{r,j}(t)}{\frac{1}{N \quad 1}_{l \quad j \quad N \quad 1} \quad DR_{il}(t)} \tag{5}$$

$E_{r,j}(t)$ → Remaining energy of node i at time t when the jth packet is transmitted by node i.

$DR_{i,j}(t)$ → Drain rate of node i at time t when the lth packet is transmitted

N → Residual energy of the node

$$N \qquad \min \; r(i) \;\; E_{cons} \quad , \;\; r(j) \;\; E_{rem} \tag{6}$$

r(i) → Residual energy of node i

r(j) → Residual energy of node j

$E_{cons}$ → Energy consumed calculated using equation (4)

$E_{rem}$ → Energy remaining

Drain rate is the ratio of difference between residual energy capacities of the node for packet l-1 and l and the difference between arrival time of these two packets.

5) Average Delay ( AvgDly ): Average RS delay( AvgDlyRS ) computes the average one-way delay observed between transmitting data from the root to each of the sinks. Average SP delay (AvgDlySP) determines the delay of transmitting packets from a source to its parent. Average delay (AvgDly ) measures the delay between transmitting data from each source to each of the sinks. It is calculated using equation (7).

$$AvgDly \quad \frac{AvgDly_{RS} \ \mathrm{Rec}_{snk} \ \ AvgDly_{SP} \ \ ATD \ \ AvgDly_{RS} \ \ \mathrm{Rec}_{src}}{\mathrm{Rec}_{snk} \ \ \mathrm{Rec}_{src}} \qquad (7)$$

$\mathrm{Rec}_{snk}$ → Amount of data packets received by all sinks

$\mathrm{Rec}_{src}$ → Amount of data packets collected by all sources

ATD → Average Tree Depth used from equation (3)

$$AvgDly_{RS} \quad \frac{RT}{r_i} \quad \frac{P_i \ \ {}^{q_i}_{j \ 1} \ \ r_j}{r_i} \quad C \, D_{ij} \ \ {}_{i,j \ path} \qquad (8)$$

$\lambda_{RT}$ → Real time data gathering rate

$r_i\mu$ → Transmission rate of real data of node i

$P_i$ → Forwarding neighbors of node i on path P

$C D_{ij}$ → Distance between node i and node j

→ Is a constant obtained by dividing a weighting constant by the speed of wireless transmission

$$AvgDly_{SP} \quad \frac{RT}{r_i} \qquad (9)$$

6) Average Packet Delivery Ratio ( AvgDR ): AvgDR measures total number of packets received .

$$AvgDR \quad \frac{Number\ of\ \ packets\ transmitted\ by\ root}{Number\ of\ \ packets\ received\ at\ the\ Sink} \qquad (10)$$

5) Simulation Results

5.1 ) Average per Source Control ( ASC ) :

Figure 1 shows the average per source controls involved in the constructions of Proposed DLMT and CMLT and the existing E-Span respectively. Our results, averaged over 15 experiments with a 95% confidence interval, have shown that the DLMT and CLMT can take up to as many as 40 times and 35 times respectively the control cost of E-Span, and this difference is expected to grow with increasing network size. The reason for such trend is due to the flooding nature of DLMT branch discovery and centralized nature of CLMT branch discovery. Since DLMT requires the eid from each source to traverse through most of other nodes and CLMT requires the selection of the bottleneck node whereas E-Span only forwards it one hop away, we do expect more control exchanges in the DLMT and CLMT model.

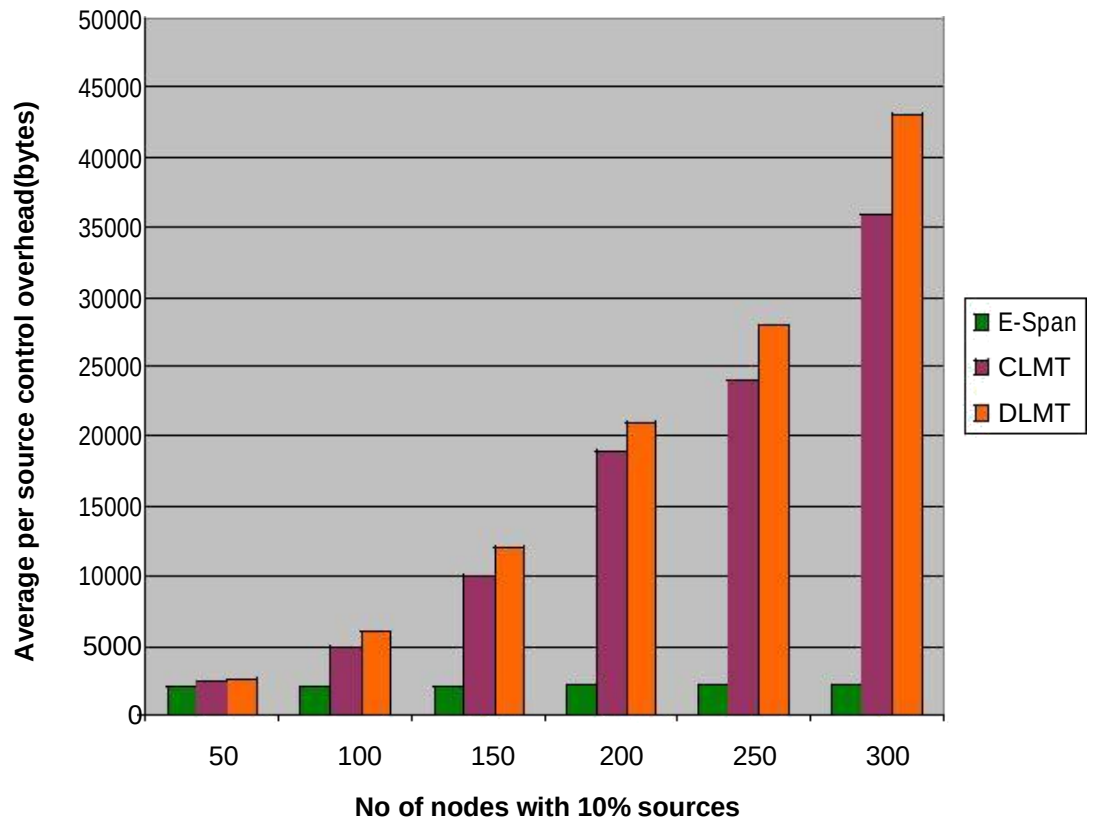

Figure 1 Average per source control

5.2 ) Average Tree Depth ( ATD ) :

Figure 2 shows the comparison of tree depth of DLMT, CLMT and E-Span as a function of network size. In fact, all the three trees are expected to grow in tree depths since a greater network size implies a greater region bounded by the sources. With radio range set at 45 meters, the root will have to traverse more hops before it can reach all the sources when this region expands. Also observe that average tree depths of DLMT and CLMT are lower than that of E-Span. Since the selection of the E-Span root is solely based on the node's energy, it is possible that this root is located at the corner of the region bounded by the sources. On the other hand, CLMT considers tree depth on the selection of the bottleneck node as the data aggregating parents. DLMT considers the tree depth in the Best Tree selection algorithm. Both the above proposed algorithms are more likely to have the tree centered at a region, therefore tree depth of both the proposed algorithms is lower than that of E-span. Results show that tree depth of DLMT is Lowest.

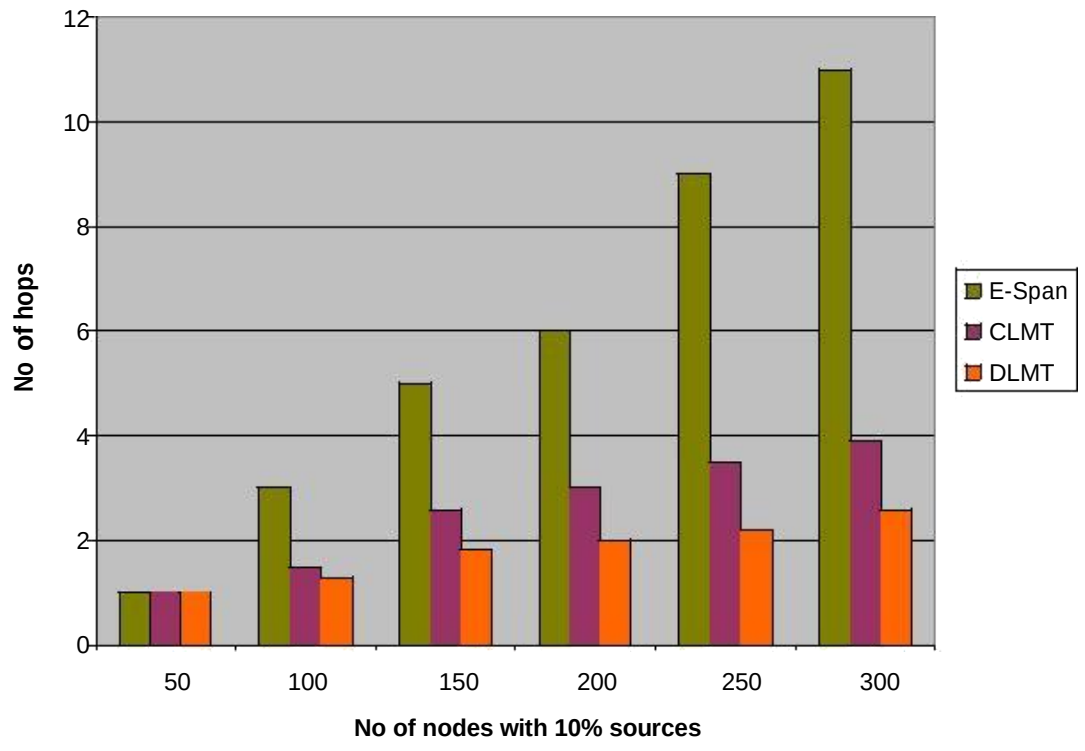

Figure 2 Comparison of tree depth overhead

5.3 ) Average Dissipated Energy( ADE ) :

To validate the impacts of data aggregation on energy savings by the use of DLMT, CLMT and E-Span, we measure the average dissipation energy and have the results averaged across 15 different experiments with a 95% confidence interval. Note that the simulation time is set at 250 seconds. Our results depicted in Figure 3 have shown a considerable amount of energy savings, approximately 35%, when data is aggregated prior to transmitting to sink using DLMT algorithm and savings approximately 27% when data is aggregated using CLMT algorithm. Such a significant saving is expected since both trees efficiently suppress the amount of traffic in the network by combining data from various sources into a single packet containing the average of all the gathered reports. We expect that this difference will continue to grow with larger network size. Also observe that DLMT has comparably equal dissipation energy as CLMT even though control exchanges are more in DLMT.

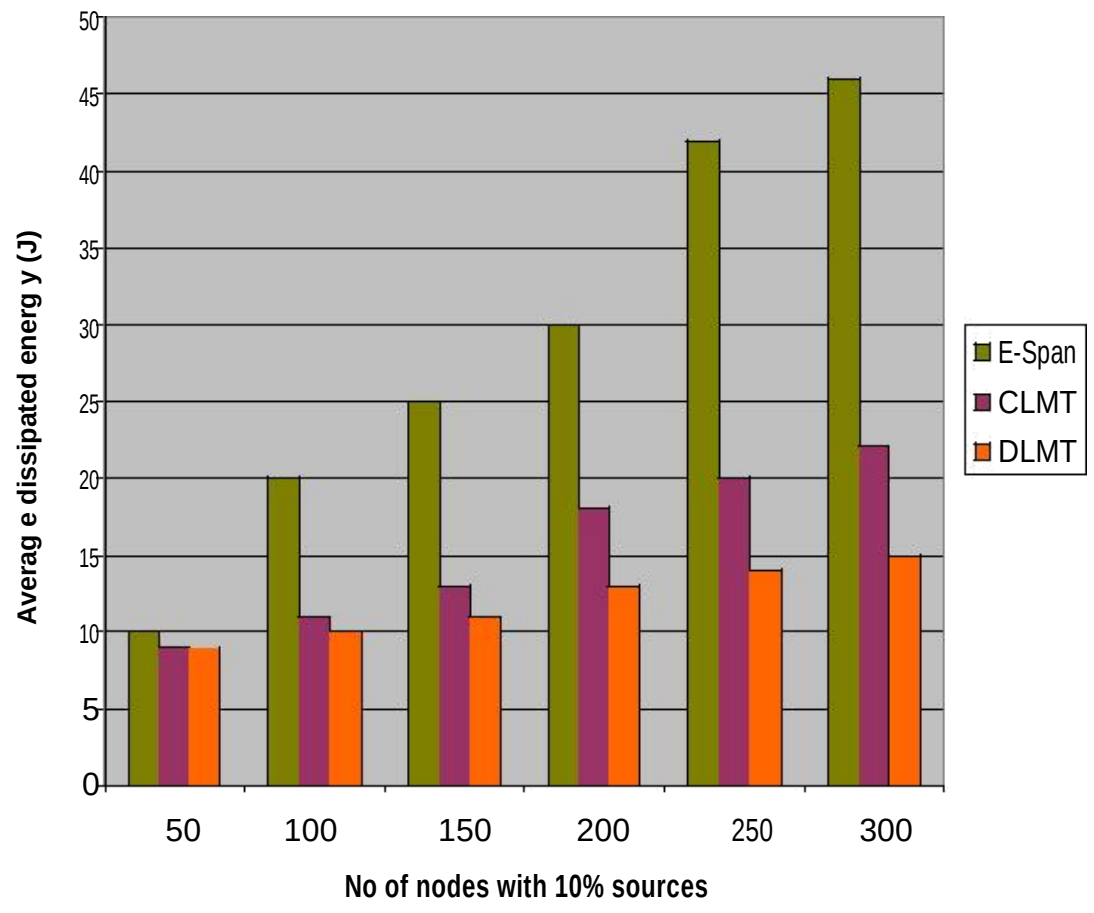

Figure 3 Average Dissipated Energy

5.4 ) Average Node Lifetime (ANLT ) :

In order to study the impact of DLMT, CLMT and E-Span on the lifetime-savings, we measure the node lifetime of each source as a function of network size for DLMT, CLMT and E -Span respectively. Each node is assigned number of with an initial energy that is randomly chosen between 12 to 18 Joules so as to limit the total simulation time at a controllable range. Since lower-energy nodes are usually being selected as leafs, they are unlikely to collect data from other sources. Given that these leafs have the same initial energy in all above schemes, the amount of lifetime-savings due to them will therefore be similar. E-Span selects the highest-energy node as the root makes this node deplete sooner than DLMT and CLMT (due to its additional duties in route coordination, exploratory data flood etc). Since the roles of the E- Span root are usually rotated among higher-energy nodes, we expect this group of nodes to have an energy dissipation rate greater than all the others. DLMT features in a way that nodes with higher energy are preferably chosen as data aggregating parents so the lifetime savings are greatest for this scheme. Both LPT and CLMT considerably extend the lifetime of each source, especially in a large network. In fact, the amount of lifetime-savings can go up to as high as 147% (As shown in figure 4 when there are 15 sources remaining, (191.4 – 77.4) / 77.4 = 147% for DLMT) and 139% (As shown in figure 4 when there are 13 sources remaining (181.6 – 75.9) / 75.9 = 139% for CLMT ) when data are aggregated through DLMT and CLMT respectively. DLMT has similar performance as CLMT in a smaller network. However, their difference starts to become more noticeable with increasing network size. In fact, our results have indicated a maximum of 13% (when there are 20 sources remaining, (215.4 – 96.1) / 96.1 – (203.2 – 96.1) / 96.1 = 13%) additional lifetime-saving when there are 20 sources remaining in the network.

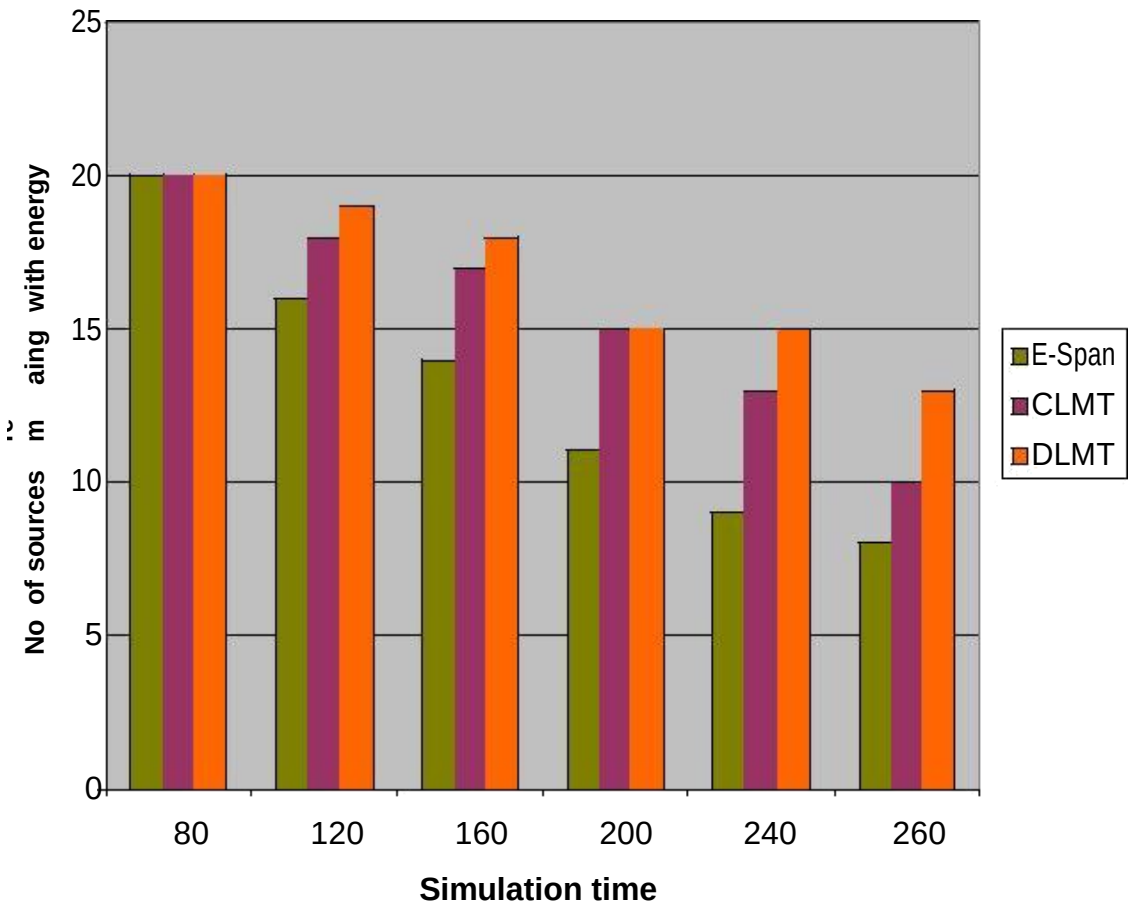

Figure 4 Average Node Lifetime

5.5 ) Average Delay ( AvgDly ):

Our next experiment compares the Average Root to sink delay observed between transmitting a compressed report at the tree root and receiving it at each sink as a function of network size for DLMT, CLMT and E-Spanrespectively. Our results, depicted in Figure 5, exhibit a trend that increases with the network size for the three schemes. As the network expands, the distance between the root and the sink increases. Consequently, the Average Root-to Sink delay also increases. Since the root selection does not depend on the positions of the 5 randomly-chosen sinks, the average distance between each root to each sink is similar for both schemes. Therefore, the difference of the delay between the three is insignificant.

To determine the delay between any pair of a source and its parent, we measure the Average Source to parent delay across 15 different experiments with a 95% confidence interval for DLMT ,CLMT and E-Span respectively. Figure 6 depicts our results. Since more participating sources increases the MAC -layer queuing delay accordingly, the Average Source to Parent delay therefore increases with network size for both schemes. Hence, the difference of the Average Source to Parent delay between the three different schemes is less predictable.

Our next experiment compares the average delay, between transmitting a data packet at each source and receiving it at each sink, for DLMT, CLMT and E-Span as a function of network size. Finally we observe from the figure 7 that DLMT and CLMT have a slightly lower delay, although quite small, than E-Span. The average tree depth for DLMT and CLMT is lower than that of E -Span, because the data in the former are only required to be forwarded for a fewer number of hops before it can arrive at the sinks.

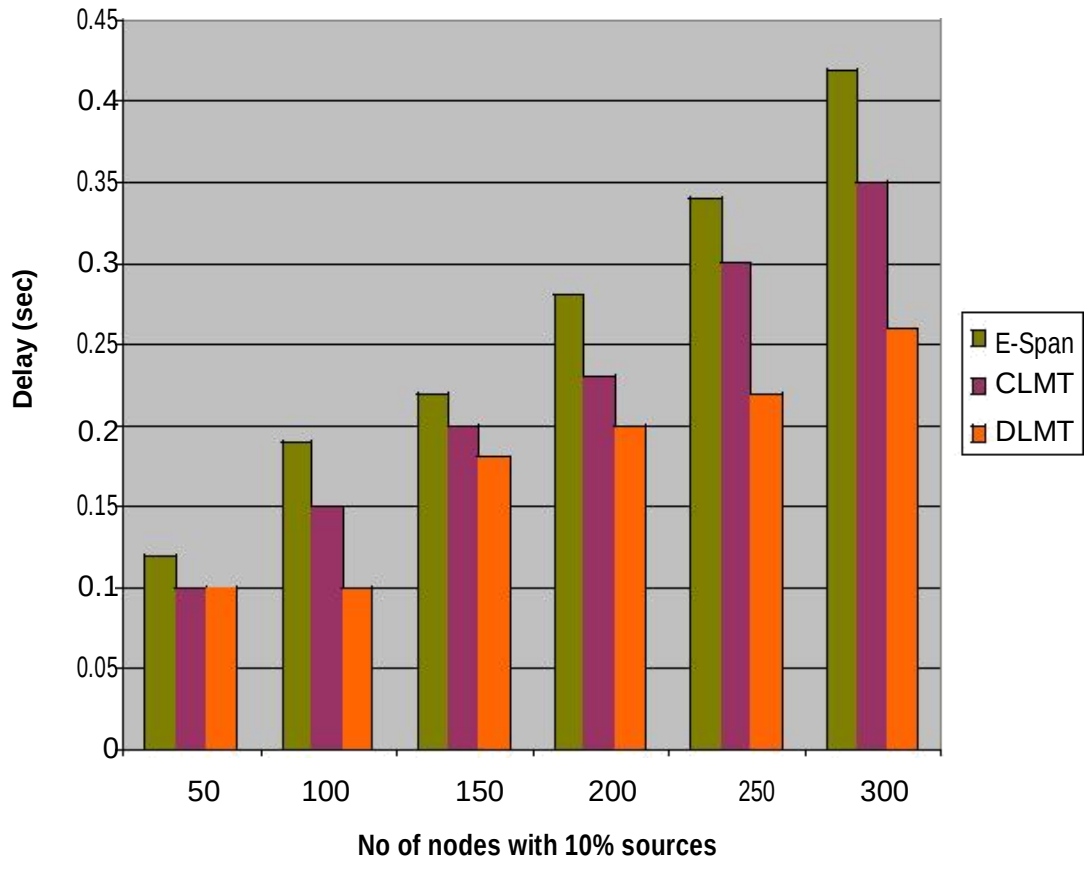

Figure 5 Average Root-to Sink delay

5.6 ) Average Packet Delivery Ratio ( AvgDR ):

Our last experiment, with its result depicted in figure 8, measures the average packet delivery ratio for DLMT, CLMT and E -Span, as a function network size, respectively. Figure 8 indicates that E-span experiences severe congestion as compared to DLMT and CLMT because in DLMT and CLMT packets are transmitted as if there is only a single source. So DLMT and CLMT are able to steadily maintain its packet delivery ratio even if network is large.

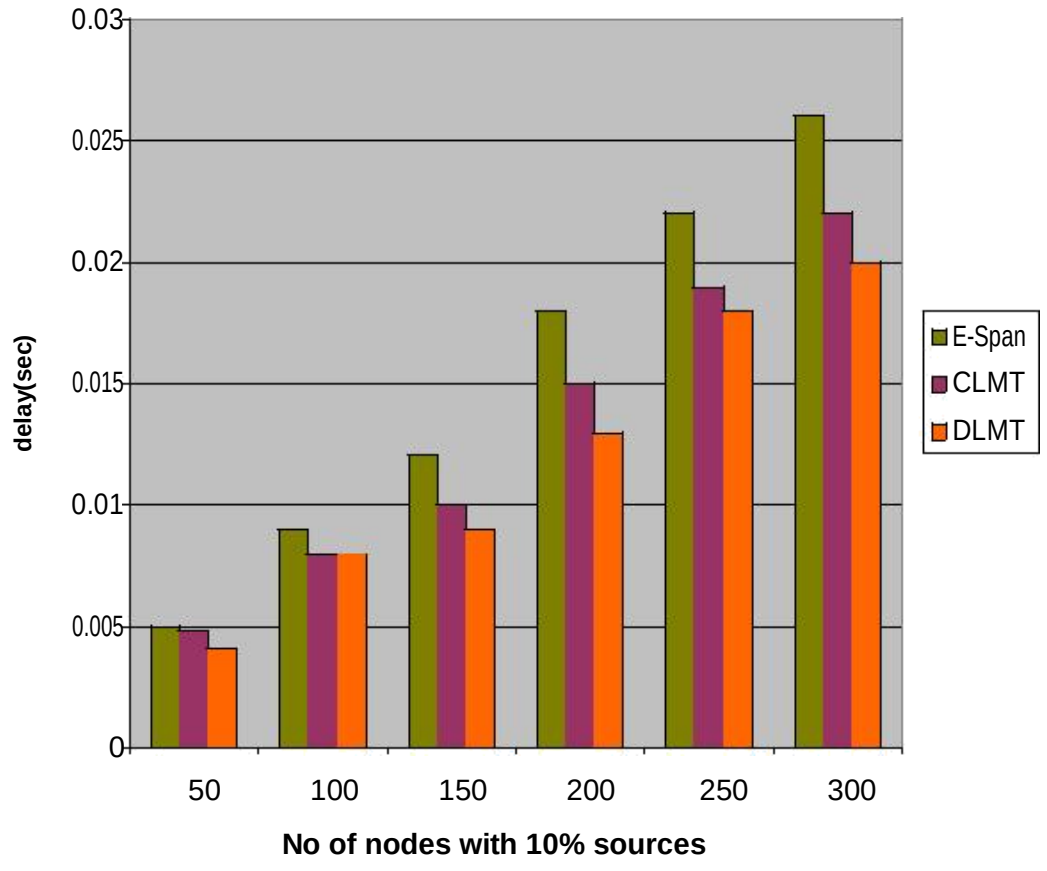

Figure 6 Average Source to parent delay

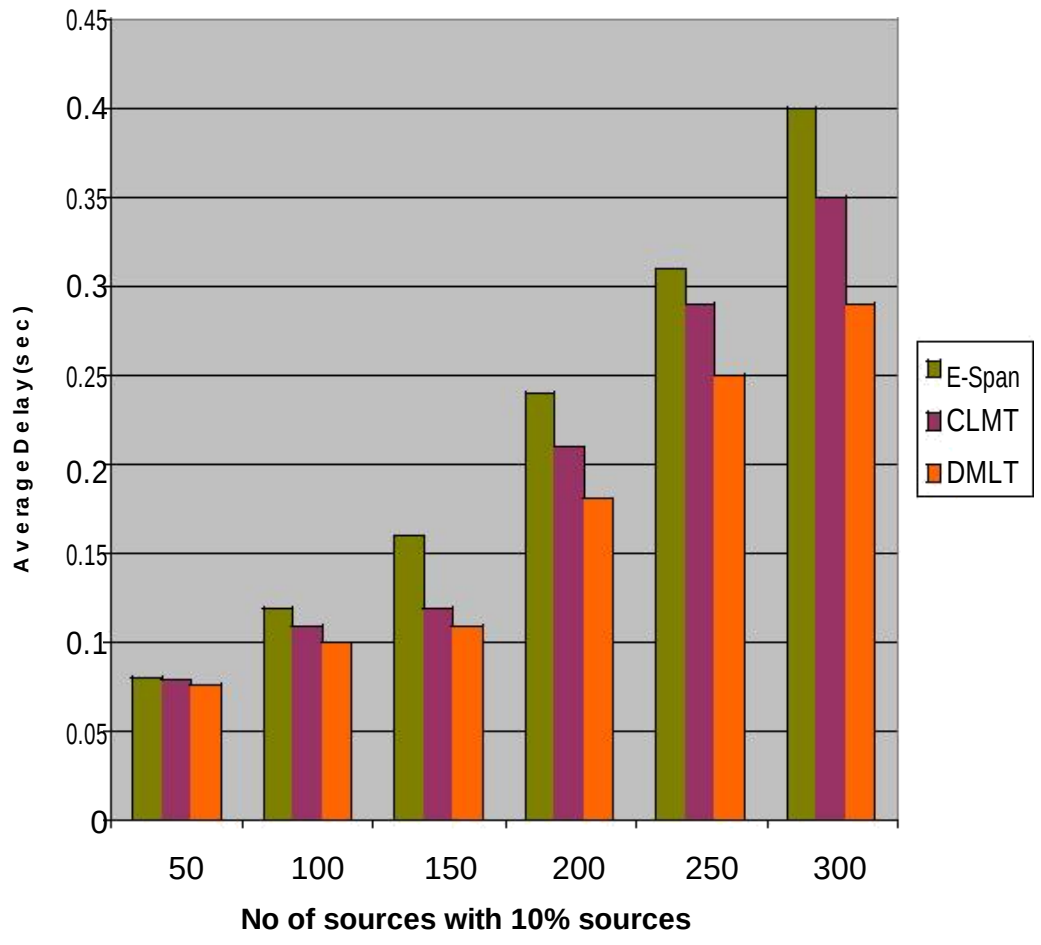

Figure 7 Average Delay

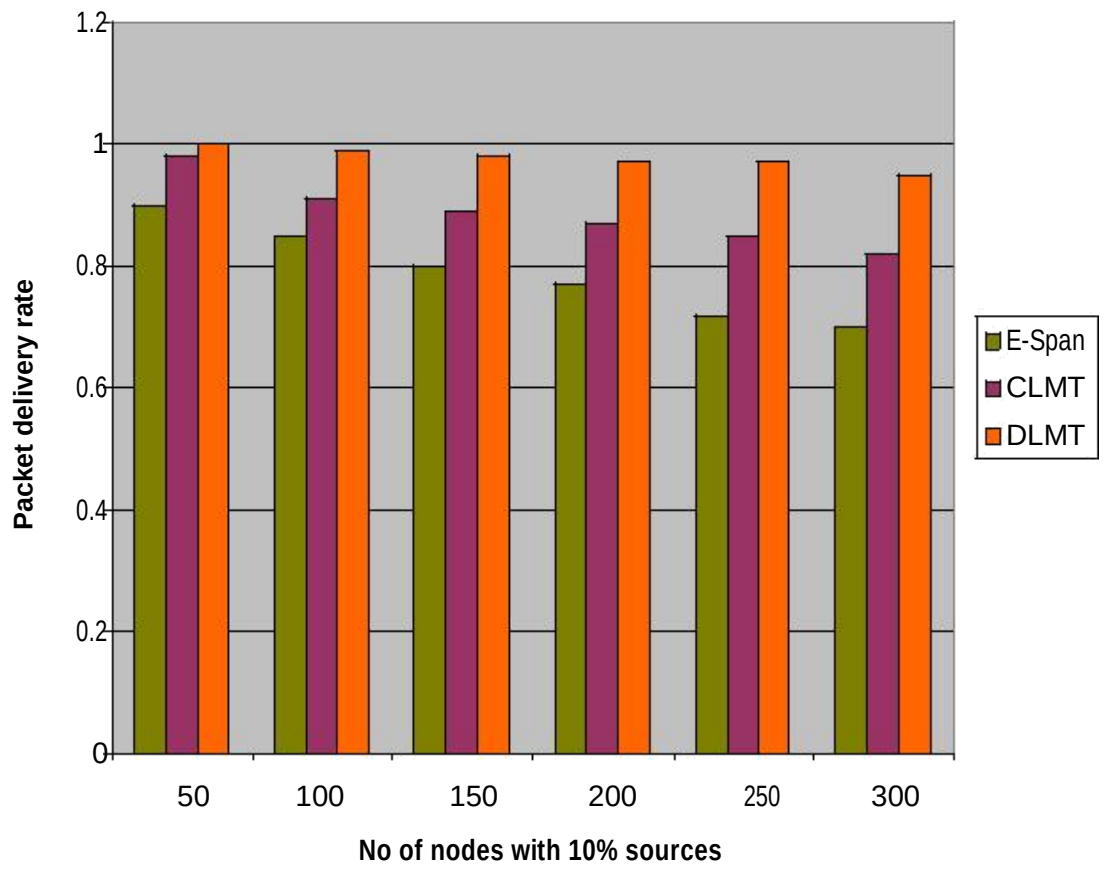

Figure 8 Average Packet Delivery Rate

## 6) Summary

This research begins with an investigation to the energy aware variant of it for their uses in data aggregation. We have demonstrated that:

> E-Span improves the design of tree construction by assigning root to be the highest energy node. Such arrangement provides root with the maximum amount of energy resources for its additional duty in coordinating the route to distant sinks. However, there is still a high chance of assigning low-energy nodes to be the data aggregating agents for the other sources.

To shorten the time and minimize the energy cost to tree reconstructions, and hence preserve the functional lifetime of all sources, we have proposed a Lifetime Maximizing Tree construction algorithm which arranges all nodes in a way that each parent will have the maximal-available energy resources to receive data from all of its children. Such arrangement extends the time to refresh the tree and lowers the amount of data lost due to a broken tree link before the tree reconstructions. We have achieved the objectives by :

1) Introducing a distributed tree construction model(DLMT) to create a tree that spans all event sources and comprises the highest tree energy using a technique similar to Reverse-Path Forwarding [7].

2) Proposing a centralized variant (CLMT) construction scheme which identifies the node that is causing a bottleneck to the set of connectivity provided by various event sources. We have simulated and compared DLMT and CLMT with and E-Span. We first validated that the tree energy of DLMT matches closely with that of the CLMT especially when there is only a few sources. We continued by comparing the amount of controls and tree depths, and have shown that DLMT is more-likely to center the tree in the middle of the area bounded by all sources. Such feature efficiently reduces the delay incurred during data collection. Moreover, our next set of results indicated a relatively steady increase of the average energy cost, delay, and packet drop rate for both DLMT and CLMT when network size increases, due to the amount of traffic suppressed by these two aggregation trees. Finally, results on average node lifetime have shown a maximum of 139 % node-lifetime extensions on the sources with the CLMT and a maximum of 147 % node-lifetime extensions with the DLMT and an additional 13% improvement when DLMT is employed instead. In fact, DLMT, CLMT and E-Span have a more pronounced difference near the tails of the two lifetime curves, implying that most of the lifetime-savings are achieved by higher-energy nodes.